\documentclass[fleqn,usenatbib]{mnras}

\usepackage{newtxtext,newtxmath}

\usepackage[T1]{fontenc}

\DeclareRobustCommand{\VAN}[3]{#2}
\let\VANthebibliography\thebibliography
\def\thebibliography{\DeclareRobustCommand{\VAN}[3]{##3}\VANthebibliography}

\usepackage{graphicx}	
\usepackage{amsmath}	
\usepackage{hyperref}
\usepackage{caption}
\usepackage{threeparttable}
\usepackage{multicol}

\newcommand{\ang}[1]{$#1\,\hbox{\rm\AA}$}

\renewcommand{\deg}{^\circ}

\newcommand{\cm}{\,\mathrm{cm}}

\newcommand{\um}{\,\mu \mathrm{m}}

\newcommand{\hour}{\,\mathrm{h}}
\newcommand{\minute}{\,\mathrm{m}}
\newcommand{\second}{\,\mathrm{s}}
\newcommand{\magnitude}{\,\mathrm{mag}}

\newcommand{\g}{\,\mathrm{g}}

\newcommand{\Gaia}{\textit{Gaia }}

\title[The primitive composition of Atira]{Nordic-based observations of binary asteroid (163693) Atira reveal a hydrated CM chondrite composition}

\author[E. M. MacLennan et al.]{
Eric M. MacLennan$^{1}$,\thanks{E-mail: eric.maclennan@helsinki.fi}
Grigori Fedorets$^{2,1}$,
Zuri Gray$^{1}$,
Teemu Willamo$^{1}$,
Tuomas Salo$^{1}$,
Anne Virkki$^{1}$,
\newauthor
Mikael Granvik$^{1,3}$,
and Karri Muinonen$^{1}$ 
\\
$^{1}$Department of Physics, University of Helsinki, PO Box 64, 00014, Finland \\
$^{2}$Finnish Centre for Astronomy with ESO, University of Turku, Vesilinnantie 5, 20014 Turun Yliopisto, Finland\\
$^{3}$Asteroid Engineering Laboratory, Lule\r{a} University of Technology, Box 848, SE-98128 Kiruna, Sweden
}

\date{Accepted XXX. Received YYY; in original form ZZZ}

\pubyear{\the\year{}}

\begin{document}
\label{firstpage}
\pagerange{\pageref{firstpage}--\pageref{lastpage}}
\maketitle

\begin{abstract}
We present photometric, spectroscopic, and polarimetric observations of (163693) Atira collected on the night of 05 March 2024. A dual-band lightcurve acquired in the $V$ and $R$ filters using the Mets\"{a}hovi Observatory (L08) in Kirkkonummi, Finland spans nearly 8~hours. From these measurements we estimate a $V-R_C = 0.43 \pm 0.07 \magnitude$. A visible spectrum and linear polarisation measurement were obtained with the Nordic Optical Telescope in La Palma, Spain. We employ a semi-empirical model to recover the spectral flux that was lost due to a mis-aligned spectroscopic slit. The corrected model reflectance spectrum indicates a primitive Cgh taxonomic type with a distinct absorption feature from 0.60 to $0.85\um$ with a depth of $\approx$7.4\%. Direct comparison with laboratory spectra of meteorites indicates close matches with CM carbonaceous chondrites. We detect a large degree of linear polarisation at a phase angle of 83$\deg$ which is consistent with (3200) Phaethon and (155140) 2005~UD, but lower than (162173) Ryugu and (152679) 1998~KU$_2$. We use a CM carbonaceous chondrite density to estimate its bulk porosity of $51 \pm 31\%$ from previously estimated bulk density. The uncertainty in bulk density and bulk porosity is dominated by the volume of the primary. Future near-infrared spectroscopic and thermal infrared observations can be used to refine the composition and bulk density. Finally, we discuss implications for ground-based physical characterisation studies of the Atira population.
\end{abstract}




\section{Introduction}

Asteroid (163693) Atira has orbital perihelion and aphelion distances of 0.502 and 0.980~au, respectively. It is the namesake and largest known object of the Atira-class of asteroids with orbits that are contained entirely within Earth's orbit (defined as having aphelia $Q < 0.983$~au). Objects with these orbital configurations rarely reach solar elongations above $70\deg$, and have been typically observed after sunset or before sunrise for only a few hours \citep{Ye_etal20,Bolin_etal22}. This poses a challenge for physical characterisation observations, in particular, for uninterrupted ground-based lightcurve observations required for rotation period and shape determination of the asteroid, which typically requires more time than is available for survey telescopes near the equator. In what follows, we avail of a rare opportunity to perform such observations from a high-latitude observatory.

Physical and compositional characterisation of inner Earth objects are essential for understanding their origin and evolution. Analysis of the known Atira population suggests a significant number of related orbital pairs \citep{delaFuente_23}. Knowledge of the spectral properties of suspected pairs can be used to confirm these relationships. Moreover, it is suspected that asteroid binary systems and orbital pairs could both be formed as a result of surface material shedding from, or the splitting of, a rapidly rotating progenitor object \citep{Walsh_etal08,Pravec_etal10,Walsh_etal12}. As the only known binary with an orbit interior to Earth, Atira is thus a high priority target for characterisation studies.

\section{Observations and Reduction}

\subsection{Optical Photometry}

We targeted Atira using the 60-cm aperture telescope at the Metsähovi Observatory (IAU observatory code: L08), owned by the University of Helsinki, on the night of 05 March 2024. The JPL Horizons\footnote{\url{http://ssd.jpl.nasa.gov/horizons}} service predicted Atira's magnitude to be $17.14 \magnitude$, which is within the estimated capabilities of a telescope aperture of this size. Sidereal tracking was used and we were able to manually track the asteroid due to its slow apparent sky motion of $1.41 \arcmin \hour^{-1}$ (per JPL Horizons). The elevation stayed above $38\deg$ (i.e., at lower culmination) for the entire night allowing for uninterrupted brightness monitoring.

\begin{table}
    \centering
    \begin{tabular}{lc}
        \hline
        R.A. & 22$^{\hour}$12$^{\minute}$ \\
        Dec. & +68$\deg$06$\arcmin$ \\
        sky motion & 1.41 $\arcmin \hour^{-1}$ \\ 
        elevation & $38\deg$ \\
        sol. elon. & $74\deg$ \\
        phase ang. & $83\deg$ \\
        eclip. lon. & $33.2\deg$ \\
        eclip. lat. & $67.7\deg$ \\
        $R_\mathit{helio}$ & 0.963 au \\
        $\Delta_\mathit{obs}$ & 0.381 au \\ \hline
    \end{tabular}
    \caption{Observing parameters of Atira during the lower transit at Metsähovi on the night of 05 March 2024.}
    \label{tab:obsgeom}
\end{table}

Mounted on the telescope was a ATIK-383L camera, having a Kodak KAF-8300 CCD chip comprised of a $3362 \times 2537$ array of $5.40\um$-sized pixels. The image has a pixel scale of $0.227 \arcsec$~px$^{-1}$ over a $12.7\arcmin \times 9.6\arcmin$ field of view. Johnson-Cousins $V$ and $R_C$ filters were used with $120 \second$ exposures and image acquisition was first carried out by alternating between filters every $\sim40 \minute$. An interleaved, alternating pattern was used after $2\hour$.

\subsubsection{CCD Noise Reduction}

Standard bias subtraction and flat field division was carried out on the science frames. After these steps a large number of warm and hot pixels were apparent and contributed a significant amount of systematic noise which we sought to mitigate. First, the sigma-clipped mean was subtracted from each science frame and a master median image was constructed using all frames in each filter separately. This resulted in a master bias frame that was scaled to the same background level and subtracted from all frames.

Next, we select a number of frames that bracket a given image in time to create a reference median-combined image. We then passed this moving ``local'' median to the \verb|ccdproc.ccdmask| algorithm from \verb|astropy| \citep{Astropy2022} which identifies and masks pixels whose values are anomalously high (i.e., hot pixels) or low relative to their local neighborhood in a $7~\mathrm{px} \times 7~\mathrm{px}$ box. The resulting bad-pixel mask is applied to the image and the process is repeated across all frames. Next, the images are binned by a factor of 3 to reduce the background noise from the sky. This binning changed the pixel scale to be $0.682 \arcsec$~px$^{-1}$. Overall, these steps resulted in an average decrease of around 80\% in the background noise of the image, as quantified by the decrease in standard deviation of the sigma-clipped background. Most of the noise reduction was from applying the spatial binning.

\subsubsection{Color \& lightcurve}

The extraction of photometry from all images was performed using the \verb|pp_photometry| function of \verb|photometrypipeline| \citep{Mommert17}. The optimal aperture in each frame was automatically chosen using a curve-of-growth analysis to maximize the signal-to-noise of the target. From this routine, instrumental magnitudes of 30 hand picked check stars were used to link across all the images to compute a continuous relative lightcurve. To provide absolute calibration across the entire dataset five stars are used, as listed in \autoref{tab:starcat}. We calculated the $V$ and $R_C$ magnitudes from \Gaia BP-RP color index using the transformations available within the Data Release 3 (DR3) online documentation \footnote{\url{https://gea.esac.esa.int/archive/documentation/GDR3/Data_processing/chap_cu5pho/cu5pho_sec_photSystem/cu5pho_ssec_photRelations.html}}. We used the calibration stars to determine magnitudes of the check stars to bootstrap a consistent calibration scheme in which all five calibration stars contribute to the zero-point of each frame, even if they fall outside the field of view.

\begin{table}
    \centering
    \begin{tabular}{lcccc}
        Calibration Star & R.A. & Dec. & $V$ (mag) & $R_C$ (mag) \\ \hline
        TYC 4463-106-1 & 22$^{\hour}$12$^{\minute}$07.2$^{\second}$ & +68$\deg$01$\arcmin$20.4$\arcsec$ & 11.65 & 11.27 \\
        TYC 4463-120-1 & 22$^{\hour}$13$^{\minute}$32.8$^{\second}$ & +68$\deg$09$\arcmin$46.9$\arcsec$ & 11.80 & 11.43 \\
        TYC 4463-1256-1 & 22$^{\hour}$14$^{\minute}$26.2$^{\second}$ & +68$\deg$09$\arcmin$57.5$\arcsec$ & 11.00 & 10.45  \\
        TYC 4463-1314-1 & 22$^{\hour}$12$^{\minute}$33.3$^{\second}$ & +68$\deg$11$\arcmin$13.7$\arcsec$ & 11.74 & 11.59 \\
        TYC 4463-412-1 & 22$^{\hour}$12$^{\minute}$59.9$^{\second}$ & +68$\deg$14$\arcmin$50.5$\arcsec$ & 11.78 & 10.99 \\
        \hline
    \end{tabular}
    \caption{Five stars used for absolute photometry calibration. The positions are provided, along with computed $V$ and $R_C$ magnitudes calculated from their \Gaia BP-RP color index.}
    \label{tab:starcat}
\end{table}

\begin{figure}
    \centering
    \includegraphics[width=\linewidth]{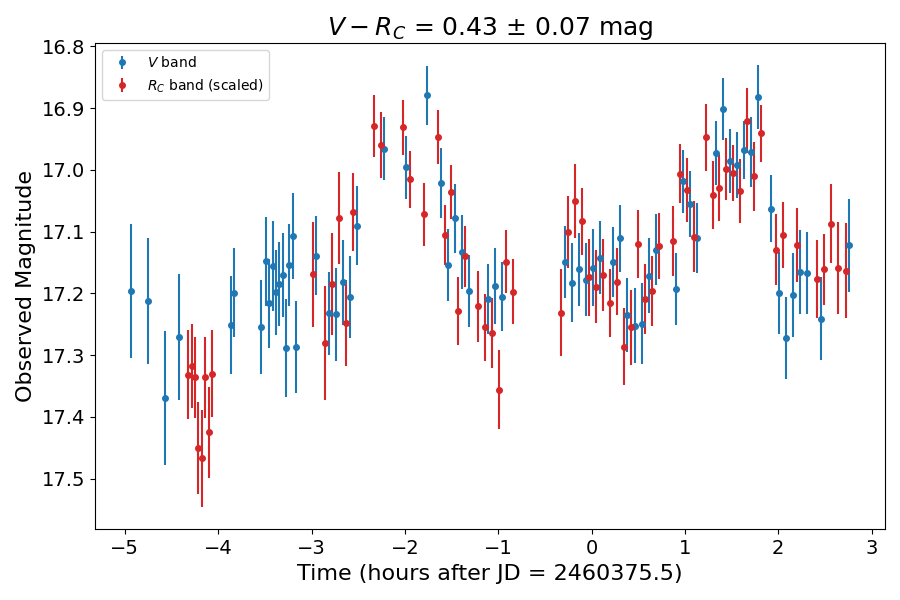}
    \caption{Observed lightcurve of Atira in $V$ and $R_C$ bands. A scaling of $+0.43 \magnitude$ is applied to all $R_C$ magnitudes.}
    \label{fig:lightcurve}
\end{figure}

The resulting apparent magnitudes for Atira are given in \autoref{app:A}. We compute the color index using measurements that were collected when images were acquired in alternating filters. A color of $V-R_C = 0.43 \pm 0.07$~mag estimated and the composite lightcurve is shown in \autoref{fig:lightcurve} after scaling the $R_C$ magnitudes.

\subsection{Polarimetry}\label{sub:polar}
Linear polarisation measurements of Atira were obtained using the ALFOSC instrument mounted on the 2.5-m Nordic Optical Telescope (NOT) on 06 March 2024, starting at UT 04:59:33.4. The polarimetric setup consists of a half-waveplate (HWP) installed in the FAPOL unit, which rotates the plane of linear polarisation, following by a calcite plate in the aperture wheel that splits the incoming light into two orthogonally polarised beams separated by $\sim15 \arcsec$ on the detector. 

The observations comprise a sequence of 100~sec exposures taken with the HWP rotated to 16 position angles between $0\deg$ and $337.5\deg$ in steps of 22.5$\deg$, obtained using the Bessel R filter (\#76 NOT filter) with $2 \times 2$ binning and non-sidereal tracking to follow the asteroid's motion. All images were bias-subtracted and flat-field corrected prior to the polarimetric analysis. Adopting the direction perpendicular to the scattering plane as the reference direction, we derived the reduced Stokes parameters $P_Q = Q/I$ and $P_U = U/I$ \citep{shurcliff1962} using the beam-swapping technique described by \citet{bagnulo2009}. We calculate the total polarisation as $P = \sqrt{P^2_Q + P^2_U}$ and the polarisation angle as $\theta = 0.5\tan^{-1}(P_U/P_Q)$. To characterise and correct for instrumental polarisation and chromatism of the HWP, we observed high- and zero-polarisation standard stars HD 251204 and HD 94851, respectively.


\begin{table}
    \centering
    \begin{tabular}{lcc}
         & {\it polarimetry} & {\it spectroscopy} \\ \hline
        UT start & 04:59:33.4 & 06:04:56 \\
        R.A. & 22$^h$13$^m$23.0$^s$ & 22$^h$13$^m$18.1$^s$ \\
        Dec. & +68$\deg$15$\arcmin$59.4$\arcsec$ & +68$\deg$17$\arcmin$41.5$\arcsec$ \\
        elevation & 19.1$\deg$ & 24.6$\deg$ \\       
        parallactic angle &  &  $-74.75 \deg$ E of N \\ \hline
    \end{tabular}
    \caption{Observation parameters of Atira at the respective start times of the NOT polarimetry and spectroscopy measurements on 06 March 2024.}
    \label{tab:NOT}
\end{table}

\subsection{Visible wavelength spectroscopy}\label{sub:spec}

Immediately following the polarimetric observations, we used the ALFOSC spectrograph to acquire three spectra of Atira starting at UT 06:04:56. The instrument setup involved a $1.3\arcsec$ slit for grism \#4 (4500 -- 9200 Å), and an exposure time of $600 \sec$ was used. Unfortunately, due to human oversight, the rotator angle of the instrument was erroneously set to $115.834\deg$ E of N and the spectroscopic slit was not aligned with the parallactic angle at the time of observation (\autoref{tab:NOT}). This resulted in gradual signal loss towards bluer wavelengths with each subsequent exposure taken. In order to account for slit loss we use the model described in the following subsection. The signal from Atira was too weak in the third exposure, but we were able to model the flux lost in the first two spectra.

The GIV solar twin HD~195034 (HIP~100963) \citep{Galarza_etal16} was observed after Atira at an airmass of 1.66, after which the flux standard Wolf 1346 \citep{Massey_etal88} was observed at an airmass of 1.73. The rotator angle of the instrument was aligned with the parallactic angle for each of these standards.

Standard spectroscopic reduction steps were performed using \verb|pypeit|. The instrument throughput sensing function was built using the spectroscopic standard and then used to flux calibrate HD~195034 and Atira. In this step an atmospheric model for La Palma included in \verb|pypeit| was used to correct for wavelength dependent extinction.

\subsubsection{Slitloss modeling}\label{sec:slitloss}

Ground--based spectroscopy obtained at non-zero zenith distances suffers from \emph{differential atmospheric refraction} (DAR), which causes a wavelength-dependent displacement of the target image along the zenith direction. In practice, this can be accounted for by aligning the slit along the direction of the dispersion, or the parallactic angle. When the spectrograph slit is not aligned with the parallactic angle, a fractional component of the chromatic displacement falls perpendicular to the slit \citep{Filippenko82}. The result is a signal loss that is wavelength-dependent distorts the apparent spectral slope, particularly for observations at low elevation.

The differential refraction $\Delta R(\lambda)$ between a wavelength $\lambda$ and a reference wavelength $\lambda_0$ is computed from the standard atmospheric dispersion relation \citep[e.g.,][]{Filippenko82}:
\begin{equation}
  \Delta R(\lambda) = \left[ n(\lambda) - n(\lambda_0) \right] \tan z ,
  \label{eq:dar}
\end{equation}
where $n(\lambda)$ is the refractive index of air and $z$ is the zenith angle of the observation. The resulting displacement projected onto the slit is
\begin{equation}
  x(\lambda) = x_0 + \Delta R(\lambda) \, \sin( \Delta{\rm PA})
               + s \, (\lambda - \lambda_0) ,
  \label{eq:xlambda}
\end{equation}
where $x_0$ is the centering offset at $\lambda_0$, $\Delta{\rm PA}$ is the angular difference between the slit position angle and the parallactic angle, and the additional linear term $s$ allows for a small empirical tilt of the spectrum across the slit, which can arise from instrumental alignment or field rotation during the exposure.

For a Gaussian point--spread function (PSF) of dispersion $\sigma$ (related to the seeing FWHM by $\mathrm{FWHM} = 2\sqrt{2\ln2}\,\sigma$) and a rectangular slit of half--width $a$, the monochromatic throughput is given by solving the integral of the normalized PSF across the slit boundaries \citep[see also][]{Szokoly2005}:
\begin{equation}
  T(\lambda) = \frac{1}{2} \left[
     \operatorname{erf}\!\left( \frac{a + x(\lambda)}{\sqrt{2}\sigma} \right)
     - \operatorname{erf}\!\left( \frac{-a + x(\lambda)}{\sqrt{2}\sigma} \right)
     \right] .
  \label{eq:throughput}
\end{equation}
This function yields the fraction of the incident flux that passes through the slit at each position, or equivalent wavelength. The observed spectrum is then modeled as:
\begin{equation}
  F_{\mathrm{obs}}(\lambda) = F_{\mathrm{true}}(\lambda)\,T(\lambda),
\end{equation}
and the corrected spectrum is assumed to be: $F_{\mathrm{corr}}(\lambda) \approx F_{\mathrm{true}}(\lambda)$.

We use tabulated refractive index differences $\Delta R(\lambda)$ for the prevailing pressure, temperature, and humidity (see \citet{Szokoly2005} for parameterization), interpolated for the airmass of the target. We set $\lambda_0 = 6200\ang$ (the approximate detector position at which the object is centered), $a = 0.7\arcsec$ (i.e., half of the slit width), 
The parameters $(x_0, s)$ are fitted via a least--squares minimization using the measured color as a constraint. The corrected asteroid spectrum is then divided by a solar analog star observed on the same night to yield the final reflectance spectrum. We note that the selection of $\lambda_0$ has an effect on the corrected flux longward of $8000~\ang$ so the spectral slope in this region carries some systematic uncertainty. On the other hand, the blue end of the spectrum ($5000-7000\ang$) is benchmarked using the $(V-R_C)_\mathrm{obs}$ color.

Because $x_0$ and $s$ are generally unknown, we empirically constrain them by minimizing the difference between a synthetic color computed from $F_{corr}$ and the independently measured color from photometric color ($(V\!-\!R_C)_{\mathrm{obs}}$) from our Mets\"{a}hovi observations. Thus, we seek to minimize the $\chi^2$ function:
\begin{equation}
  \chi^2(x_0,s) = \left[(V\!-\!R_C)_{\mathrm{model}}(x_0,s) -
                       (V\!-\!R_C)_{\mathrm{obs}}\right]^2 ,
\end{equation}
where $(V\!-\!R_C)_{\mathrm{model}}$ is obtained from performing synthetic photometry on a reference Solar spectrum \citep[\texttt{sun\_reference\_stis\_002}][from the CALSPEC library]{Colina_etal96} multiplied by the asteroid's reflectance values.  This semi--empirical calibration allows the wavelength dependence of slit losses to be corrected even when the centering offset ($x_0$) and second order linear chromatic effects (captured by $s$) are uncertain. Given the uncertainty in $V-R_C$ and the dependency on the slitloss modeling, we run our model for different values of $(V\!-\!R_C)_{\mathrm{obs}}$. Uncertainties in the synthetic color are estimated from adjusting the individual uncertainties of the spectral data points via several thousand Monte Carlo trials.


\begin{figure}
    \centering
    \includegraphics[width=0.95\linewidth]{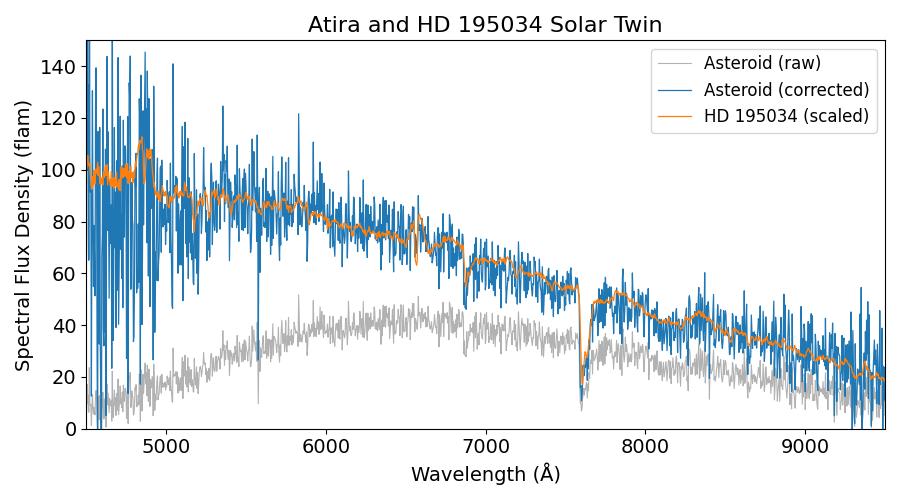}\\
    \includegraphics[width=0.95\linewidth]{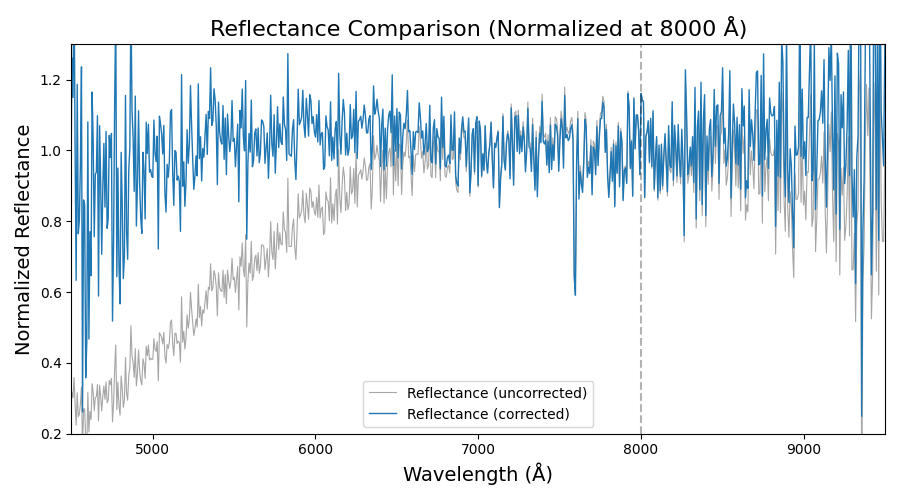}
    \caption{Spectral fluxes and spectral reflectances for the raw (uncorrected) and corrected (via slitloss modeling) of Atira. A scaled flux spectrum of the Solar twin HD~195034 used to compute the reflectance is also shown. Asteroid spectra are binned by a factor of 2 for easier visual comparison.}
    \label{fig:specloss}
\end{figure}

\section{Results and analysis}

\subsection{Lightcurve}

\begin{figure}
    \centering
    \includegraphics[width=0.99\linewidth]{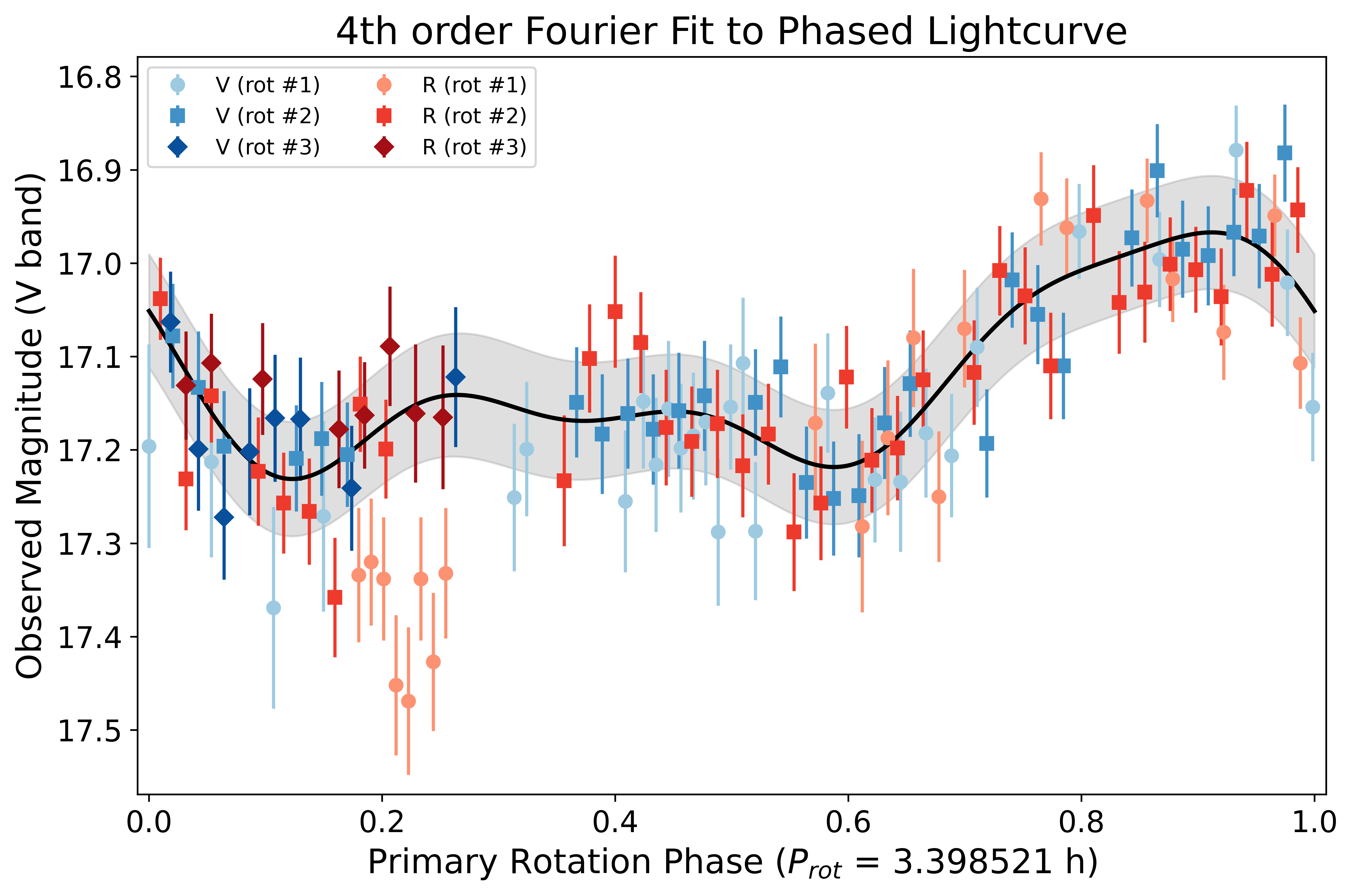}
    \caption{Lightcurve observations phased to the primary rotation period of the primary. Symbols signify the filter and rotation number counted from the first observation. A fourth-order Fourier function is fit to points excluding an inferred mutual event observed in the $R$ filter during the first rotation (light red circles). The shaded gray area represents the 1-$\sigma$ prediction range. See text for details.}
    \label{fig:phasedlc}
\end{figure}

The lightcurve duration extends to just short of 8 hours, and spans nearly three rotations of the primary. We show the rotational phases (taking the first observation time as an arbitrary zero point) for each photometry point using the rotation period of the primary, $3.39852\hour$ \citep{Deleon_etal24} in \autoref{fig:phasedlc}. The symbol shapes refer to the rotation number and colors referencing the filter. The brightness variations are found to be consistent across several rotations, except for points in the $R_C$ filter for the first rotation phase of 0.2. We interpret this drop in brightness to be a mutual event of the secondary being occulted by the primary or the shadow of the secondary passing across the surface of the primary.

We fit a fourth-order Fourier function to points acquired in both filters, excluding the aforementioned possible mutual event. The fitted function is shown as a black curve in \autoref{fig:phasedlc} and has an amplitude of around $0.25 \magnitude$. The prediction interval is shown as a gray area and represents the uncertainty in obtaining a single new measurement. The prediction interval is consistent across all rotation phases at around $0.05\magnitude$, which is also on the order of the magnitude uncertainty in the individual measurements. The short drop in brightness at the start of the observations exceeds this prediction interval, supporting our suggestions that it is a valid detection of a mutual event.

\subsection{Degree of polarisation}

We measure Atira's reduced Stokes parameters as $P_Q = (31.78 \pm 1.93) \%$ and $P_U = (-0.74 \pm 1.96)\%$ at a phase angle of $\alpha = 83^{\circ}$. The corresponding total polarisation is $P \sim P_Q$ and the polarisation angle is $\theta = (-0.66 \pm 1.76)^{\circ}$ relative to the direction perpendicular to the scattering plane, indicating that the polarisation vector is aligned, within uncertainties, with the expected scattering geometry. 

\begin{figure}
    \centering
    \includegraphics[width=\linewidth]{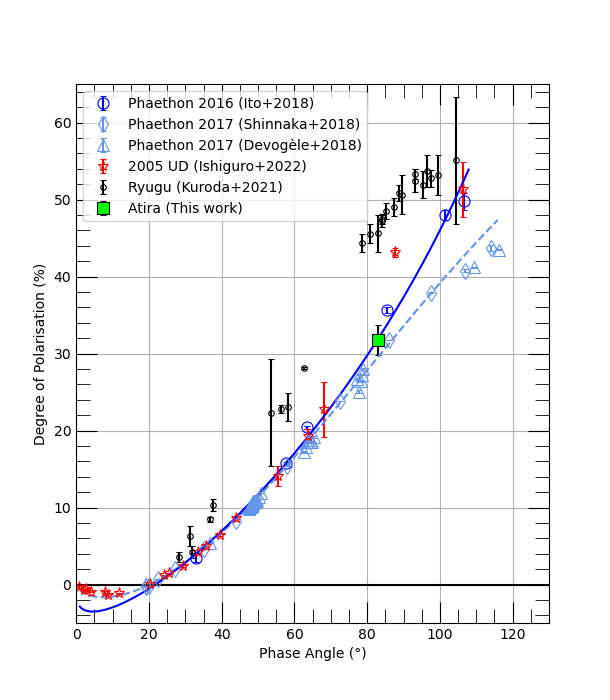}
    \caption{Degree of linear polarisation of Atira in comparison with other low-albedo NEAs. The solid and dashed curves represent the modelling polarimetric phase curve fitted to the 2016 and 2017 observing epochs, respectively. }
    \label{fig:polar}
\end{figure}

In Figure \ref{fig:polar}, we compare our Atira measurement with polarimetric data of other low- to moderate-albedo NEAs: (3200) Phaethon \citep{ito2018, shinnaka2018,devogele2018}, (155140) 2005 UD \citep{ishiguro2022}, and (162173) Ryugu \citep{kuroda2021}. A notable feature in the comparison sample is the significant discrepancy in the maximum polarisation $P_{\rm{max}}$ of Phaethon between the 2016 and 2017 observing campaigns. To emphasise this difference, we model the polarimetric phase curves of the two epochs separately using the empirical-trigonometric relation of \citet{lumme_muinonen1993}:
\begin{equation}
P(\alpha) = b\ \sin^{c_1}(\alpha)\ \cos^{c_2}\Big(\frac{\alpha}{2}\Big)\ \sin(\alpha - \alpha_0)
\end{equation}
where $b, c_1,$ and $c_2$ are free parameters that shape the curve, $\alpha$ is the phase angle, and $\alpha_0$ is the inversion angle -- the phase angle at which the polarisation switches sign. The resulting fits indicate $P_{\rm{max}} \sim 50\%$ at $ \sim 107^{\circ}$ for the 2016 epoch \citep{ito2018}, and $P_{\rm{max}} \sim 45\%$ at $ \sim 130^{\circ}$ for the 2017 epoch \citep{devogele2018}. These variations have been attributed to differences in sub-observer latitude during the respective observing geometries, which suggests latitude-dependent regolith properties possibly driven by asymmetric solar heating at perihelion. We refer the reader to \citet{MacLennan_etal22} for further reading on the topic.

In the context of asteroid polarimetry, Umov's law describes the inverse correlation between $P_{\rm{max}}$ value and the geometric albedo of the scattering body. This means that the polarisation of light scattered by a low albedo surface will be stronger than that scattered by a high albedo surface. This trend is clearly visible in \autoref{fig:polar}. With a geometric albedo of $p_v = 0.040-0.045$ \citep{sugita2019,tatsumi2020}, Ryugu exhibits the strongest polarisation among the set of comparison objects. In contrast, with $p_v = 0.088-0.109$ \citep{Devogele_etal20,ishiguro2022}, 2005~UD shows a somewhat lower polarisation, similar to the higher $P_{max}$ dataset from 2016 \citep{ito2018}. Published albedo estimates for Phaethon span a relatively broad range, $p_v = 0.09-0.14$ \citep{hanus2016,ito2018,shinnaka2018}, consistent with its overall polarimetric behaviour. Although insufficient for precise albedo determination, the measured polarisation at $\alpha = 83^{\circ}$ clearly indicates the Atira has a dark surface, with a higher albedo than Ryugu ($>4.5\%$) and more similar to Phaethon.

\subsection{Spectral properties}\label{sub:specprop}

\begin{figure}
    \centering
    \includegraphics[width=\linewidth]{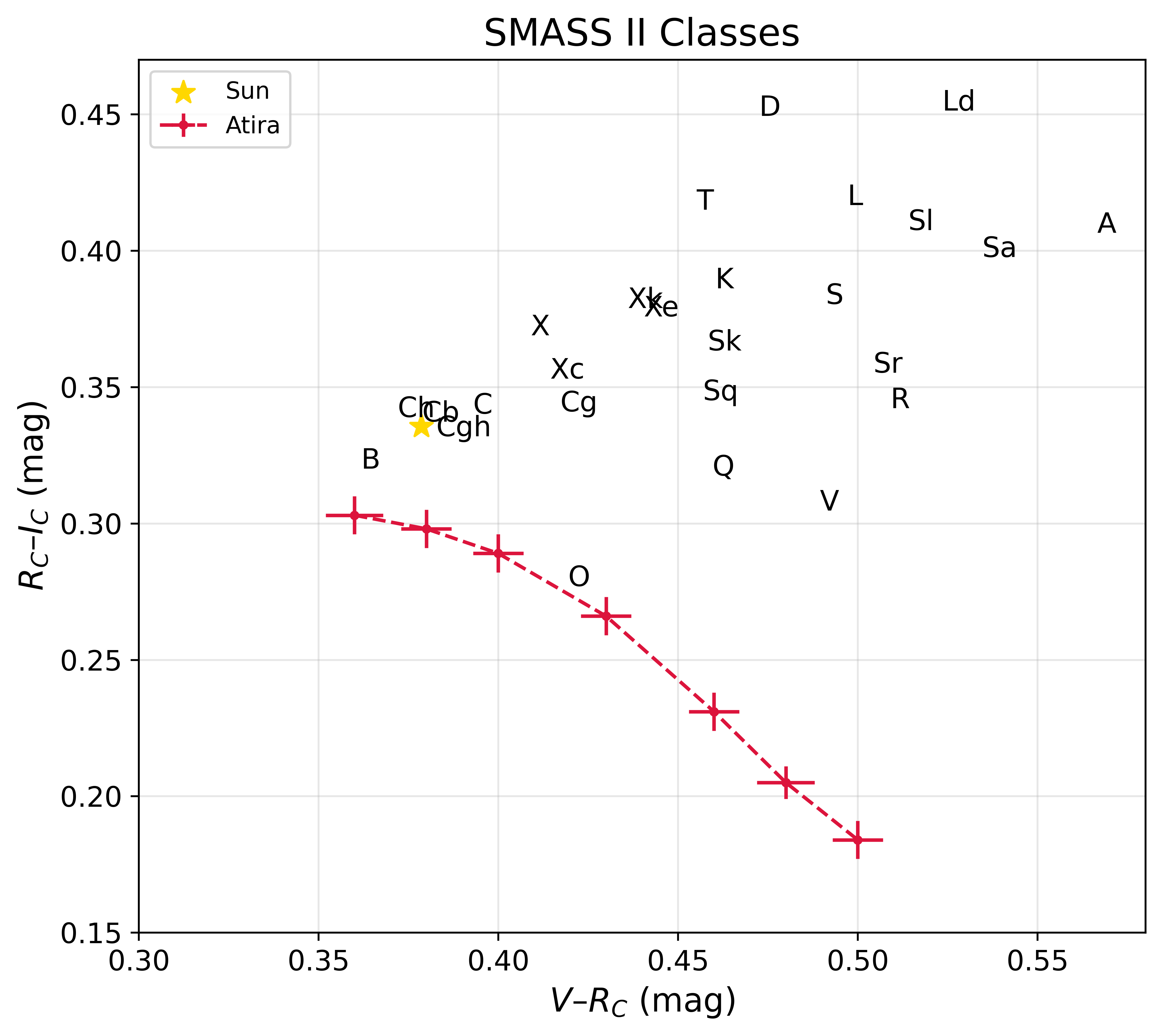}\\
    \includegraphics[width=0.98\linewidth]{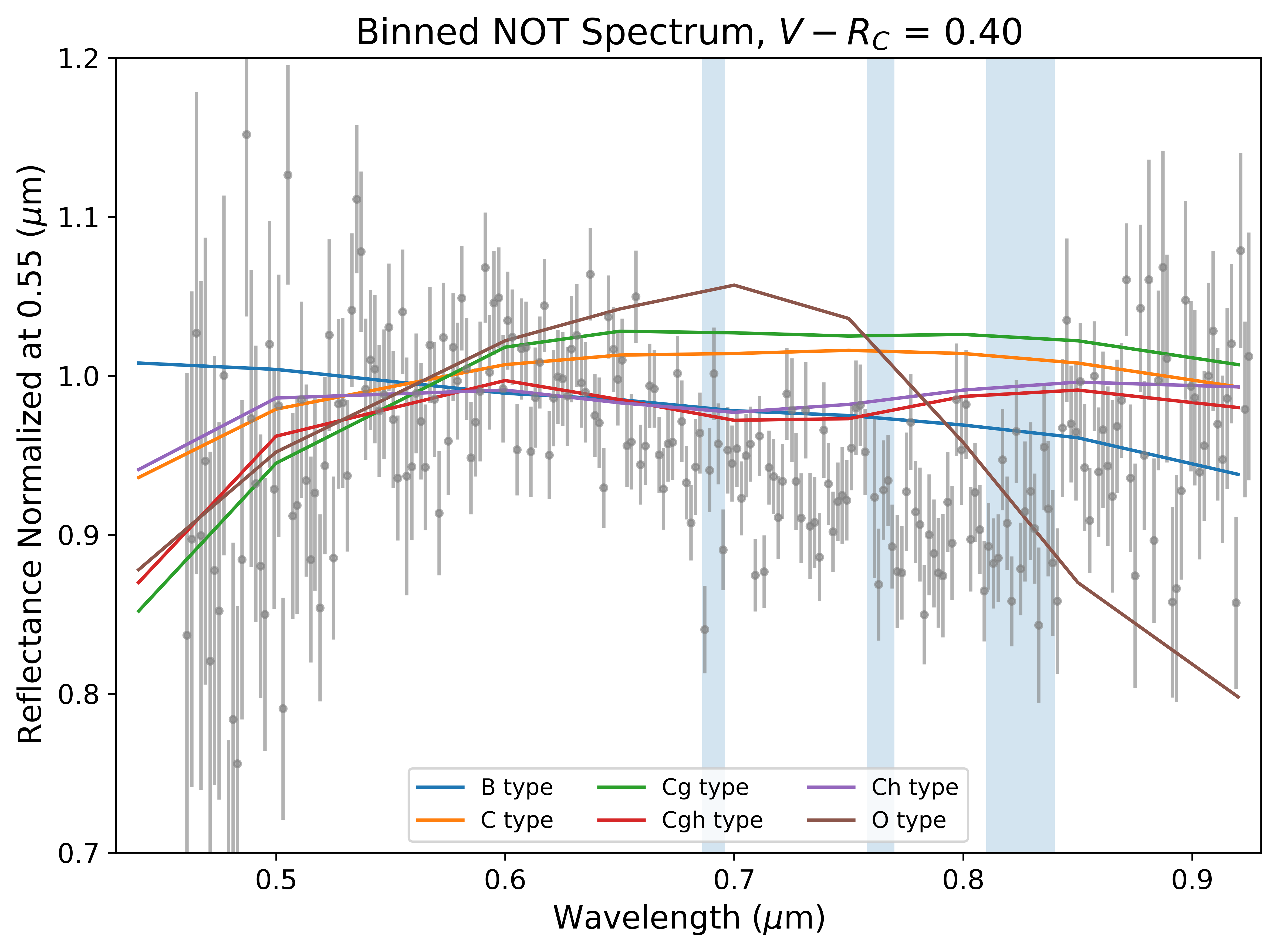}\\
    \caption{Comparison to Bus spectral types. Top: Color plot showing the colors from Atira's spectrum (red points connected by a dashed line) compared to mean spectra of SMASS II taxonomic types. The yellow star shows the colors of the Sun and blue point is (3200) Phaethon. Synthetic colors from Atira's corrected spectrum for different input $V-R_C$ colors is shown by connected red points. Bottom: Atira's spectrum has been binned by a factor of 20 and compared to mean spectra of six SMASS II classes. The gray area shows a telluric feature that affects the reflectances.}
    \label{fig:taxcolors}
\end{figure}

We determined a color of $V-R_C \approx 0.43$ from our Mets\"{a}hovi lightcurve observations, with a large uncertainty of 0.07~mag. We explore the effect of changing the $V-R_C$ color input to the slitloss model. To encapsulate and visualize the effect of this we calculate synthetic $R_C-I_C$ colors as a function of the input $V-R_C$. The color-color plot in the top panel of \autoref{fig:taxcolors} shows the synthetic colors of the NOT spectrum in comparison to the colors of the SMASS II taxonomic classes \citep{Bus+Binzel02}. The $V-R_C$ and $R_C-I_C$ colors are inversely related and have smaller $R_C-I_C$ values than all the taxonomic classes. This results is due to the drop in reflectance at wavelengths common to the $I_C$ band (0.7$-0.9\um$), as seen in the bottom panel of \autoref{fig:taxcolors}, and which we discuss further below.

In the bottom panel of \autoref{fig:taxcolors}, the wavelengths that are affected by a telluric feature are shown as shaded blue regions. Intriguingly, the  $V-R_C \approx 0.43$ and $R_C-I_C \approx 0.26$ colors are proximal to the rare O-type class. We rule this out based on the dissimilar spectral behaviour beyond 0.6 $\um$. Other classes such as the C, Cgh, Ch, and B types are close to the range of possible colors in the top panel of \autoref{fig:taxcolors}. The steep UV dropoff seen in Atira's spectrum is not consistent with the B type template. Similarly, the dropoff in reflectance for C and Ch types at wavelengths shorter than $0.60\um$ does not match Atira's spectrum. The descriptions of Cg and Cgh classes in \citet{Bus+Binzel02} mention a strong dropoff shortward of $0.55 \um$, which is most consistent with Atira.

The reflectance values in the binned spectrum (\autoref{fig:taxcolors}) show a clear absorption feature spanning wavelengths $0.65-0.85\um$ which is a characteristic of the Cgh class. This feature can be seen in the bottom panel of \autoref{fig:specloss} but is more apparent after binning and is notably stronger than the model Cgh spectrum in \autoref{fig:taxcolors}. The strength of this feature explains why Atira's $R_C - I_C$ color is notably lower by about 0.04~mag. An absorption feature centered at 0.7$\um$ is often attributed to the Fe$^{2+} \rightarrow $Fe$^{3+}$ in phyllosilicate minerals, which are products of aqueous alteration \citep{Vilas+Gaffey89}. 

\begin{figure}
    \centering
    \includegraphics[width=\linewidth]{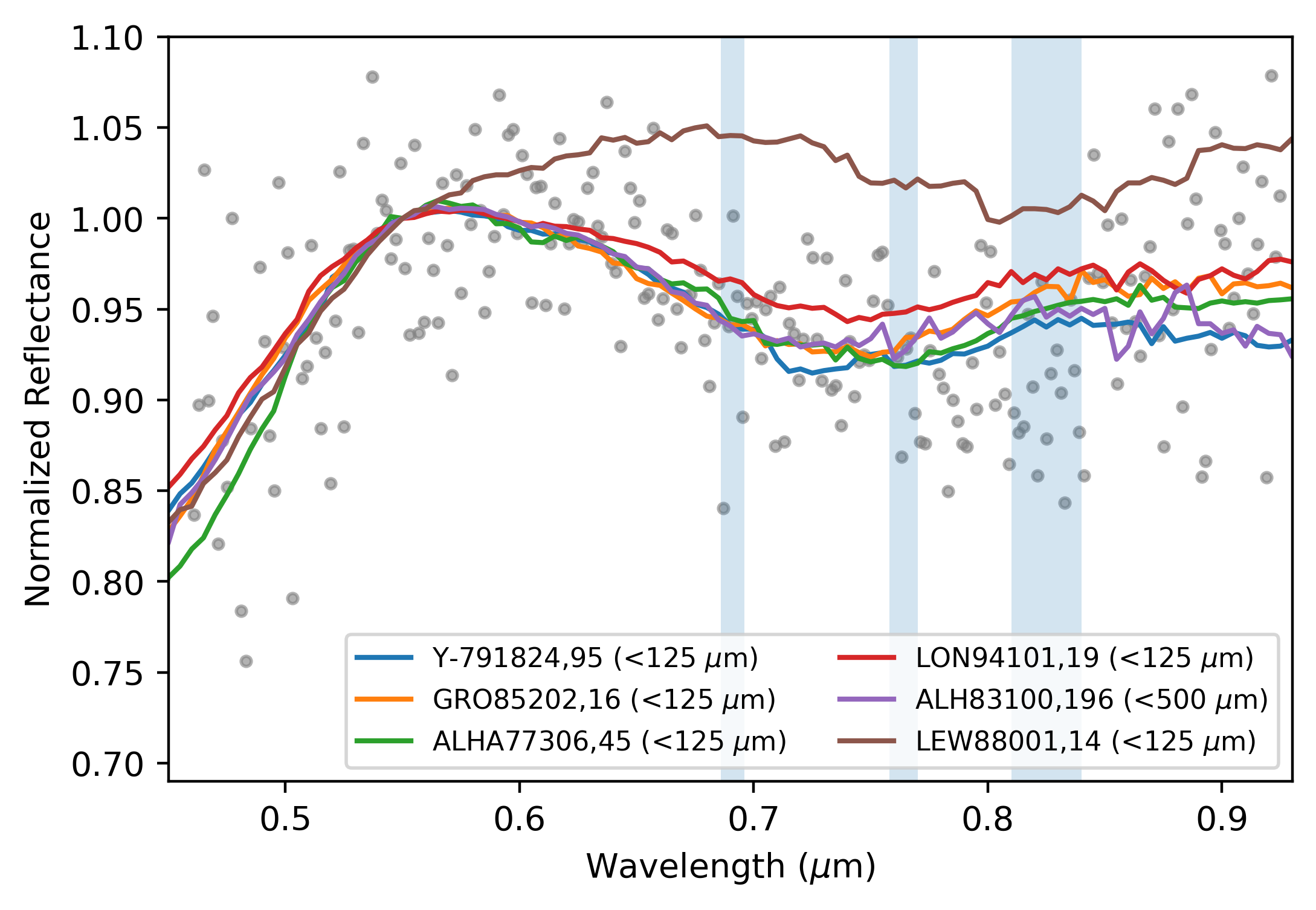}
    \includegraphics[width=\linewidth]{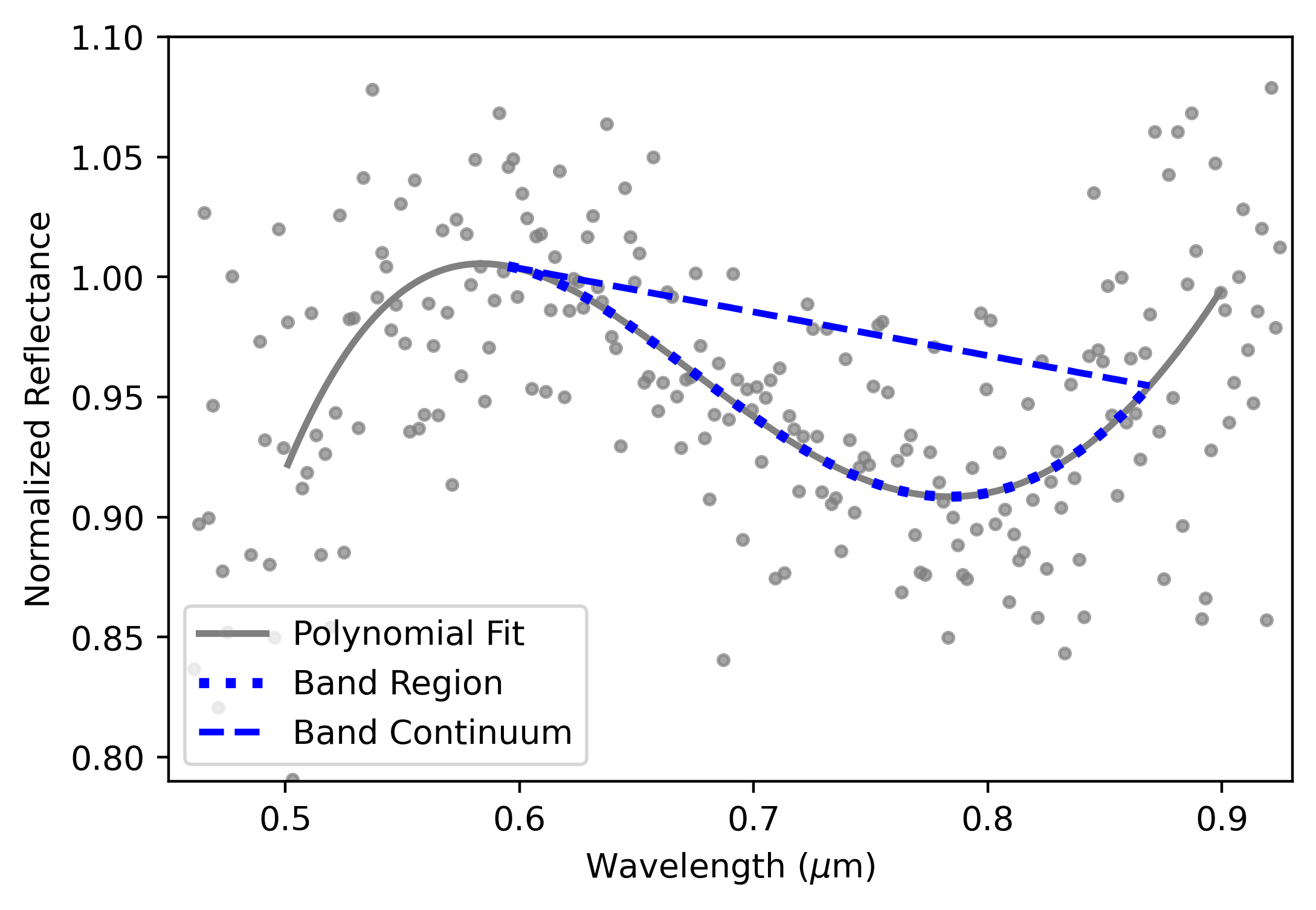}\\
    \includegraphics[width=\linewidth,clip,trim=0 0 0 1cm]{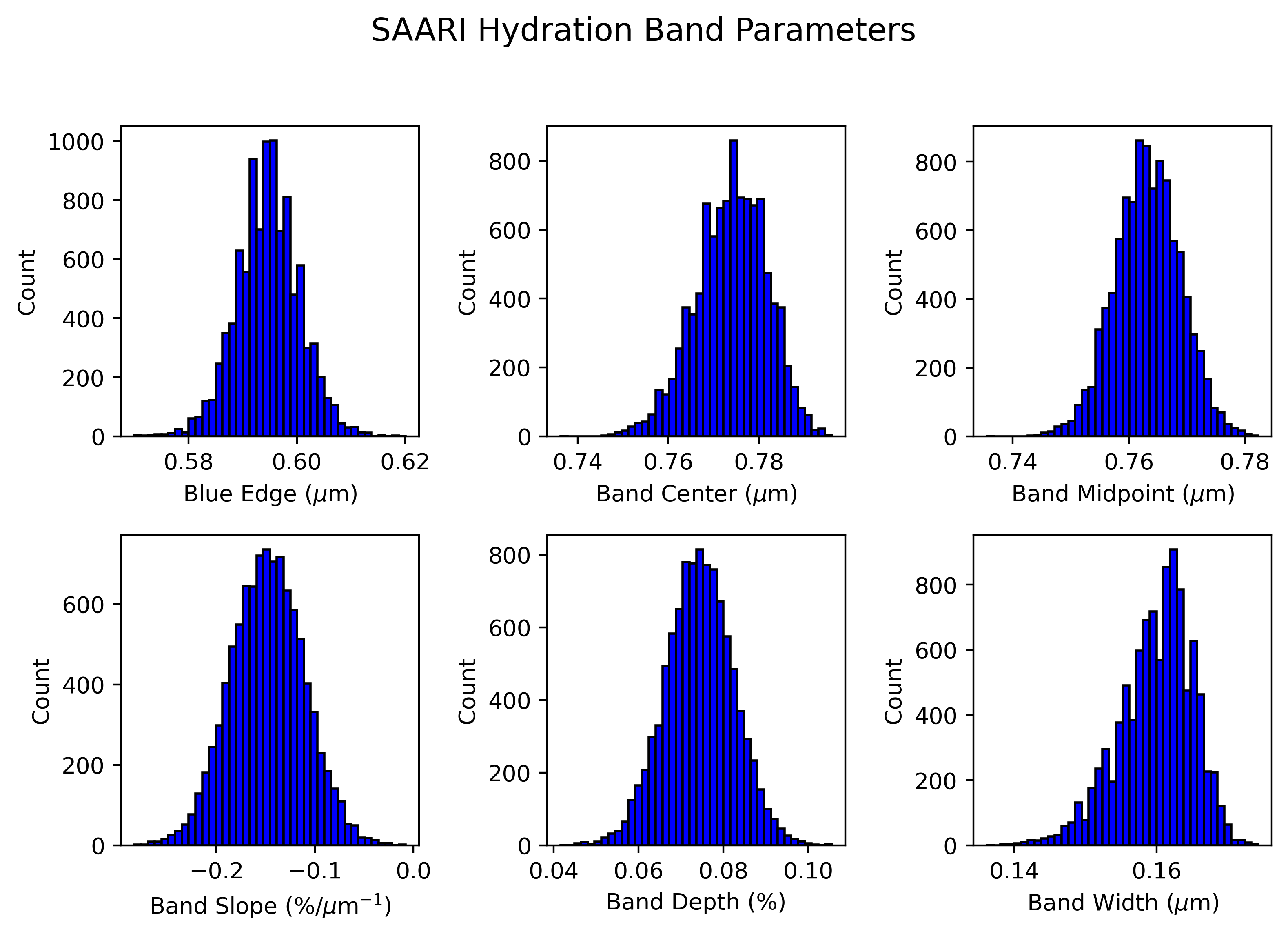}\\
    \caption{Top: Four CM meteorites plotted alongside the binned NOT spectrum of Atira. Telluric absorptions are shown as blue regions. Middle: Fitted 4$^{th}$ degree polynomial and definition of the 0.7 $\um$ band region and continuum used in the SAARI routine. Bottom: Band parameter distributions from 10,000 Monte Carlo trials. }
    \label{fig:hydration}
\end{figure}

We perform a search for meteorite analogs in the RELAB spectral database\footnote{https://sites.brown.edu/relab/} using the M4AST online tool \citep{Popescu_etal12}. All of the top matches unambiguously consisted of CM (Mighei-like) carbonaceous chondrites: GRO85202,16 ALHA77306,45 Y-791824,95 LON94101,19 and ALH83100,196. As shown in \autoref{fig:hydration}, Atira's reflectance values appear to be highly consistent with these meteorite spectra throughout the entire sampled wavelength range. In particular, the steep dropoff in reflectance shortward of 0.55$\um$ and the depth of the 0.7$\um$ absorption feature, which have both been noted to indicate extensive aqueous alteration \citep{Hiroi_etal96,Tatsumi_etal23}. The strong absorption seen in Atira's spectrum suggests a high degree of aqueous alteration \citep{Fornasier_etal14}, which we discuss further below. The 0.7$\um$ band has not been seen in any CI chondrites, which may likely be connected to the higher abundance of Fe-rich serpentine-group phyllosilicates in the CMs \citep{Cloutis_etal11}. We include the LEW~88001, a highly weathered CM meteorite \citep{MarlowMason1990}, in \autoref{fig:hydration} because it exhibits an absorption centered near 0.83$\um$, where Atira exhibits lower reflectance in comparison to the other CM spectra. We discuss this topic further in \autoref{sub:origins}.

To analyse the 0.7$\um$ feature we use a modified version of the SAARI (Spectral Analysis of Asteroids for Reflectance Investigation) band analysis routine \citep{MacLennan_etal24}. SAARI typically applies the \verb|loess| smoothing algorithm to spectral reflectance data and calculates band positions, depth, area, and width after a linear continuum is defined. In this version of SAARI, and partly following \citep{Morate_etal18}, we use a 4$^{th}$ order polynomial fit to reflectance in the wavelength range 0.5-0.9$\um$ to define the band region. The decision to fit a polynomial instead of a \verb|loess| function is due to the relatively low signal to noise of the data and the presence of telluric features--particularly near 0.82$\um$ that may complicate the \verb|loess| smoothing approach. Due to the noise in Atira's spectrum and the neutral/negative slope of the absorption feature it is difficult to empirically identify its red edge from the reflectance data. Therefore, we fix the location of red edge of the feature at $0.87\um$ to match the shoulder of the feature seen among CM meteorites (top panel of \autoref{fig:hydration}). A linear continuum is then constructed across the band from the red edge to the tangent point to the polynomial fit near the maximum reflectance around 0.6$\um$ (i.e., the blue edge). From these parameters, we run 10,000 Monte-Carlo trials to estimate the band parameters and their uncertainties as shown by the distributions in the bottom panel of \autoref{fig:hydration}. The band center (minimum reflectance) is estimated to be 0.774 $\pm$ 0.008 $\um$ and the depth to be 7.4 $\pm$ 0.8\%. The band midpoint--defined as the center wavelength of the band at the half-minimum--is 0.763 $\pm$ 0.006 $\um$ which, when compared to the band minimum, indicates some slight asymmetry in the absorption feature. The band slope, estimated at -0.148 $\pm$ 0.037 $\% \um^{-1}$, is simply the slope of the band continuum.

Finally, we can use Atira's spectral classification to infer an albedo range. The mean albedo of Cgh asteroids to be within the 5-7\% range as calculated by \citet{Lantz_etal13} (42 asteroids from the IRAS catalog) and \citet{Burbine_etal24} (8 objects from the NEOWISE catalog). This range is notably larger than the low albedo of $p_v \approx 0.02$ previously reported in the literature for Atira \citep{rondon2022,Deleon_etal24}. Such estimations of Atira's albedo have been done using the absolute magnitude diameter relationship, which we note the inherent and significant uncertainty in the former due to the high phase angles that Atira is observed at. Based on the mean Cgh albedo range and noting Atira's polarisation in proximity to both Phaethon and 2005~UD, we infer that its true albedo may instead lie between 5 and 11\%.

\section{Discussion}

\subsection{Binary nature \& origins}\label{sub:origins}

After phasing the Metsähovi lightcurve to the rotation period of the primary, we are able to identify a mutual event as a drop in brightness of 0.2~mag compared to the expected lightcurve. This brightness change was only seen in the $R_C$ filter, and its magnitude is unlikely explained by surface color variation. Theoretically, multi-filter observations before and during an occultation or eclipse of the secondary by the primary could be used to search for color difference of the components. From our observations alone it is impossible to deduce if this mutual event is an occultation of the secondary by the primary or an eclipse of the primary on the secondary. Given the large time gap between our observations and the last set of lightcurves collected in 2019, combined with the uncertainty in orbital elements of the secondary, extrapolating the position to our observations is feasible. Additional lightcurves are needed to better constrain the orbital parameters of the secondary from the work of \citet{Deleon_etal24} to model these data.

The reflectance spectrum of Atira is a Cgh-type in the SMASS taxonomic system, which indicates a primitive composition. Primitive asteroids in the wider binary population are found to be rare \citep{Minker+Carry23}, but some still been identified using radar observations and spectroscopically studied. For example, 1996~FG$_3$ is another binary asteroid that is likely to be primitive based on its low albedo and spectral characteristics \citep{deLeon_etal11}, and which has been studied because of its prior status as a primary mission target. The ultra-blue near-infrared spectrum \citep{Perna_etal14} of the triple system (153591) 2001 SN$_{263}$ \citep{Becker_etal15} indicates a B-type taxonomy and no 0.7$\um$ feature was detected in its spectrum. As of writing, we are unaware of any other binary or multiple system that has a 0.7$\um$ phyllosilicate feature.

In a literature search we found three near-Earth asteroid with reported Cgh classifications: (276049) 2002~CE$_{26}$, (385186) 1994~AW$_1$ \citep{Popescu_etal19}, and (365246) 2009~NE \citep{Somers_etal10,Binzel_etal15}, although the spectrum of 2009~NE is not publicly unavailable \citep{Somers_etal10}. \citet{Popescu_etal19} calculated the band depth of 1994 AW$_1$ as larger ($\sim$3.7\%) and more red-shifted (0.76$\um$) compared to 2002 CE$_{26}$ (1.6\% and 0.71$\um$). To our knowledge, Atira has the deepest 0.7$\um$ absorption detected among the asteroid population. 

Main-belt Cgh asteroids, of which there are dozens, have been linked to CM carbonaceous chondrites and may represent parent body interiors that were then exposed via impacts \citep{Vernazza_etal16}. Comparison of Atira's spectral properties (\autoref{sub:specprop}) to primitive Main Belt families studied in the PRIMASS survey \citep{Morate_etal16,Morate_etal18,Morate_etal19} can give insight into its origins. We calculate the slope of a linear fit to the reflectances from 0.55 to 0.9~$\um$ to be -1.83 $\pm$ 0.37~\%~1000\ \AA$^{-1}$, which aligns with the $1\sigma$ lower limit of the size-slope distribution for primitive Main Belt families \citep{Morate_etal19}. The band depth of Atira's phyllosilicate feature is remarkably larger than the dozens of Cgh-type asteroids, with almost all of them having depths in the 1-4\% range \citep{Fornasier_etal14,Morate_etal18,Morate_etal19}, and the position of this feature is shifted to longer wavelengths compared to virtually all the asteroids in the PRIMASS survey. Interestingly, we found two Cgh asteroids, 57400 (in the Sulamitis family) and 106919 (Clarissa family), that have both an unusually deep ($\sim6\%$ absorption depth) and red-shifted (0.74 $\um$) phyllosilicate feature \citep{Morate_etal18}. No overall correlation between band depth or band positions and taxonomic classes was found in the PRIMASS studies. On the other hand, \citet{Fornasier_etal14} showed that the 0.7$\um$ band positions of CM chondrites are shifted to longer wavelengths (0.71–0.75 $\um$) compared to hydrated asteroids. This possibly indicates different mineral abundances and a compositional mismatch between the meteorite collection and the main-belt population, and/or grain size differences. Given that Atira's absorption feature is deeper and redder than the asteroids and meteorite samples, we propose that compositional and grain size properties are more similar to CM meteorite powders than Cgh asteroid regolith.
Variations in the 0.7$\um$ band are common, with some objects showing rotational variation \citep{Rivkin_etal15}.

The bulk density of Atira's primary has been estimated to be $1.43 \pm 0.87 \g \cm^{-3}$ \citep{Deleon_etal24}. From this constraint, and taking a grain density of $2.80 \g \cm^{-3}$ representative of CM samples \citep{Macke_etal11}, we estimate a bulk porosity of $51 \pm 31 \%$. The relatively large uncertainty is chiefly driven by the uncertainty in the size ($D_\mathit{eff} = 4.92 \pm 0.95$~km), and thus the volume, of Atira. Future work could reduce this uncertainty with more precise determination of the diameter, from thermal infrared observations and modeling, for example.

The \citet{Greenstreet_etal12} model found that asteroids on orbits close to Atira have a 55\% and 22\% likelihood of being delivered via the $\nu_6$ and 3:1 resonances, respectively, giving two possible escape mechanisms from the Main Belt \citep{Bottke_etal02}. Within the orbital evolution model of \citet{Granvik2018} the following probabilities of main-belt source regions/resonances are provided for Atira: 0.65 $\pm$ 0.03 ($\nu_6$ secular resonance), 0.19 $\pm$ 0.02 (Hungaria region), 0.12 $\pm$ 0.03 (3:1 MMR), 0.03 $\pm$ 0.01 (5:2 MMR), and 0.007 $\pm$ 0.002 (Phocaea region). Given it's retrograde spin axis \citep{Deleon_etal24}, the Yarkovsky effect on the Atira system causes its orbital semimajor axis to decrease \citep{Bottke_etal02b}. As a consequence, the orbital evolution of Atira is most likely to have escaped from a larger semimajor axis \citep{Vokrouhlicky2015}. For example, Atira would have originated from an family in the inner Main Belt for it to have encountered the $\nu_6$ resonance, or from the middle Main Belt into the 3:1 resonance. Given the high rarity of C-complex asteroids in the Hungaria family and background \citep{Lucas_etal17,Lucas_etal19}, we rule the Hungaria region as a possible source for Atira.


\subsection{Thermal environment \& heating effects}

Assuming thermal equilibrium, Atira's daytime surface temperatures can reach up to 530~K (260$\deg$~C) at its perihelion distance of 0.502~au, which falls in the range of temperatures categorized as stage I heating of carbonaceous chondrites in the classification of \citet{Nakamura05}. In stage I--defined as temperatures below 300$\deg$~C--temperatures are not large enough to completely dehydrate phyllosilicates, as found in laboratory analyses of naturally-heated CM samples \citep{King_etal21b}. Thermal decomposition of minerals are predicted to occur on Phaethon as it experiences stage III-IV heating \citep{MacLennan+Granvik24}. Compared to Phaethon, whose surface experiences perihelion temperatures of 1050~K ($\approx\!780\deg$C) \citep{MacLennan_etal21}, Atira has experienced a lesser degree of heating. Disappearance of the phyllosilicate absorption feature in the spectrum was observed when the Murchison meteorite was experimentally heated to a temperature of $400\deg$~C \citep{Vilas+Sykes96}. Thus, the presence of this feature suggests that Atira's surface has not experienced this degree of heating. In various laboratory heating experiments, the 0.7$\um$ feature remains and phyllosilicates in CMs remain hydrated up to 300$\deg$C \citep{Hiroi_etal96,King_etal21a}.

Alteration of accessory minerals, which are not tracked in the heating stages of \citet{Nakamura05}, have also been studied in the laboratory. For example, \citet{Garrene_etal14} showed that the dehydroxylation of ferrihydrite and goethite found in CM samples occurs at around $225\deg$~C and $250\deg$~C, respectively. Comparison between the meteorites (top panel of \autoref{fig:hydration}) the reflectance values and Atira's spectrum appear lower near 0.8$\um$. This may be an observational artifact, or could indicate the presence of iron oxides/hydroxides alteration minerals \citep{Vilas_etal94}. For example, the CM sample LEW 88001 exhibits a minimum near $0.83\um$, which \citet{Cloutis_etal11} suspects could be a "composite of the 0.7-$\um$ serpentine band and the iron oxide/hydroxide band", the latter of which is located around $0.9\um$. In the case of LEW~88001, these ferrous oxyhydroxides (a.k.a. rust) are a weathering product due to terrestrial contamination \citep{MarlowMason1990,Lee_etal23}. Given that Atira's surface temperatures are high enough to vaporize any loosely-bound adsorbed water in the subsurface, which we suppose could theoretically oxidize the oxyhydrides near the surface causing them to undergo some weathering.

Our attention is drawn to the laboratory work of \citet{Opeil_etal20} which shows that CM meteorites exhibit a unique negative thermal expansion coefficient around 235~K in which the material experiences expansion--as opposed to contraction--with decreasing temperature. The authors attribute this phenomenon to the contraction experienced among phyllosilicate/serpentine layers that dominate CMs \citep{King_etal17}. Nominally, thermal gradients within larger rocks and boulders cause mechanical stress gradients that exceed the material strength, resulting in fracturing, also called thermal fatigue \citep{Delbo_etal14,Molaro_etal17}. However, non-homogeneous expansion and contraction would act to increase internal stress of a material without the need for particularly steep thermal gradients \citep{Molaro_etal17}. The negative expansion transition behavior for CMs occurs in the temperature range of 210-240~K, which falls in the expected diurnal temperature range for Atira. Recent laboratory experimentation shows the increased susceptibility of CM meteorites to repeated heating over a wide range of temperatures \citep{Latisa_etal2026}. We therefore predict that Atira's surface is relatively deficient of boulders and larger rocks (cm-scale), resulting in a low thermal inertia compared to similarly-sized primitive asteroids \citep{MacLennan+Emery21}.


We also compare Atira's properties to that of the near-Sun asteroid (1566) Icarus, which has also been imaged by radar and experiences extreme temperatures during its perihelion passage at 0.186~au. Icarus is a Sq-type asteroid with a very low ($\sim$2\%) radar albedo -- which is defined as the ratio of the object’s radar cross section and its projected area -- that suggests a porous near-surface regolith \citep{Greenberg_etal17}, which is consistent with its low estimated thermal inertia \citep{Novakovic_etal24,MacLennan_inprep}. Whereas Icarus' radar albedo lies at the low extreme of the NEA population, Atira's radar albedo of $0.41 \pm 0.10$ suggests a dense, or low-porosity, regolith. The relatively large uncertainty in Atira's albedo is derived from the uncertainty in its size, but is clearly elevated relative to the general NEA population detected by radar \citep{Virkki_etal22}. We note that high radar albedos are typically observed only for M-type (metal-rich) asteroids (which in this case conflicts with the spectra) or icy bodies (which is impossible for asteroids orbiting as near the Sun as Atira). Furthermore, as noted by \citet{Deleon_etal24}, the radar backscatter of Atira is strongly specular, which we note was seen also for Icarus \citep{Greenberg_etal17}. Strongly specular radar backscatter typically indicates fine-grained regolith, where it is likely that thermal processes have disintegrated majority of cm-scale and larger regolith particles. In contrast, for example (101955) Bennu -- whose surface is known to be depleted of fine-grained regolith and dominated by cm-scale and larger particles -- has a very diffuse radar-backscatter profile \citep{Nolan_etal2013,Lauretta_etal2019}. Abundant cm-scale particles could also increase the radar albedo if their permittivity is sufficiently high, but this possibility for Atira is diminished by the specularity indicating mostly fine-grained regolith. Atira's circular polarisation ratio of $0.21 \pm 0.01$ represents a typical value relative to the NEA population \citep{Virkki_etal22} but is slightly larger than B-type NEAs Phaethon, Bennu, and 2001 SN$_{263}$, which have mean value of $0.18 \pm 0.03$ \citep{Taylor_etal19}. On the other hand Icarus' value lies above this range. In this case, the circular polarisation ratio reveals little of the surface composition. 

\subsection{Implications for characterizing Earth-interior objects}

The steady-state population of Atira objects is estimated to be in the hundreds \citep{Greenstreet_etal12}, with some stable regions at high orbital inclinations and low semimajor axes \citep{Ribeiro_etal16}. Less than 50 have been discovered, and physical characterisation has only been carried out for a few objects \citep{rondon2022}. Atira was discovered by the Lincoln Near-Earth Asteroid Research (LINEAR) survey \citep{Stokes_etal00} operating in the New Mexico desert (latitude: $+33.818\deg$). A number of all-sky surveys including the DECam at the Cerro-Tololo Inter-American Observatory (latitude: $-30.169\deg$), the ZTF \citep{Ye_etal20,Bolin_etal22} at Palomar Observatory (latitude: $+33.356\deg$), \citep{Sheppard_etal22}, and the LSST at the Rubin Observatory (latitude: -$30.245\deg$) operate during astronomical twilight. Combining data across these surveys may be useful for studying the physical characteristics of the population.

Another promising and forthcoming source of observations comes from the Vera C. Rubin Observatory's Legacy Survey of Space and Time \citep[LSST, ][]{ivezic2019}, which will be a transformative astronomical survey that will increase the discovery rate by several times across all small Solar System populations. Although the majority of LSST's time will be devoted to the southern-sky wide-fast-deep survey, an additional Northern Ecliptic Spur survey will enhance the number of the discoveries throughout the ecliptic. Furthermore, additional observations aiming at finding inner-Earth objects have been proposed. The LSST near-Sun twilight survey has been originally proposed by \citet{seaman2018}. As the twilight time is short, in order for more robust orbit determination it was suggested to follow a more conventional four-observation strategy for the twilight survey (contrary to the pairs used in the main Wide-Fast-Deep survey). 

The near-Sun Twilight Micro Survey has been accepted for the first year of LSST \citep{scoc2023} as the only micro-survey\footnote{\url{https://survey-strategy.lsst.io/baseline/micros.html} Micro-surveys are programmes that are operated outside the main Wide-Fast-Deep survey pattern, while not exceeding one percent of the total observing time.}. However, the first year of the twilight survey will be ran in a truncated testing mode -- only during morning twilight on every fourth night. The observations will follow the original plan, with data taken in three ($riz$) filters, with exposures lasting 15 s, and four repeated visits. \citet{scoc2023} suggest that dedication of more time to the twilight survey in the subsequent years of LSST will be made based on the results of the first-year survey.

The details of the yield from the LSST twilight survey have been investigated by \citet{schwamb2023}. Summarising their findings, assuming the NEO model of \citet{Granvik2018}, LSST is expected to discover on the order of tens of percents of Atira- and 'Aylo'chaxnim-type objects down to a brightness of $V = 16$~mag. LSST will act as a discovery survey for inner-Earth asteroids. However, their subsequent self-characterisation by LSST is not foreseen, as it is not for near-Earth asteroids in general \citep{kurlander2025}. In particular, the optimal potential for near-polar region observatories to complement LSST discoveries of inner-Earth objects lies in obtaining dense long lightcurves for asteroid shape determination.

Depending on the time of the year, observatories at high latitudes experience longer twilight hours compared to locations near the equator. Thus, these locations are better optimised for observing asteroids with lower solar elongation. In some cases when the orbital inclination is sufficiently large, a large enough declination may allow an object to be observed through an the entire night. Atira and other Earth-interior asteroids are accessible from ground-based observatories exclusively at high phase angles and thus absolute magnitudes from photometric phase curves are difficult to estimate. Conversely, polarimetric measurements at high phase angles are valuable for discerning the albedo and distinguishing taxonomic types. 

\begin{figure}
    \centering
    \includegraphics[width=1\linewidth]{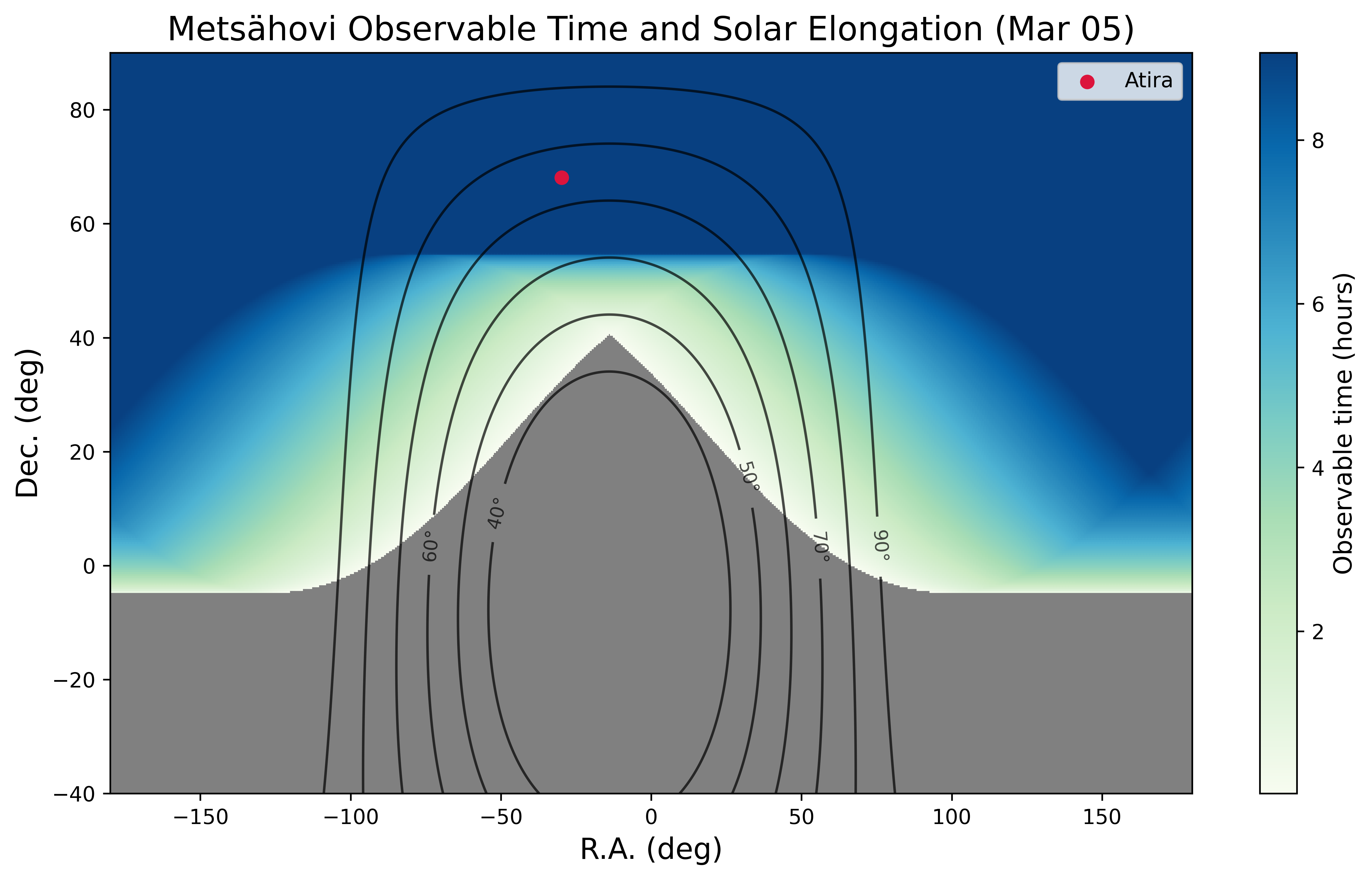} \\
    \includegraphics[width=1\linewidth]{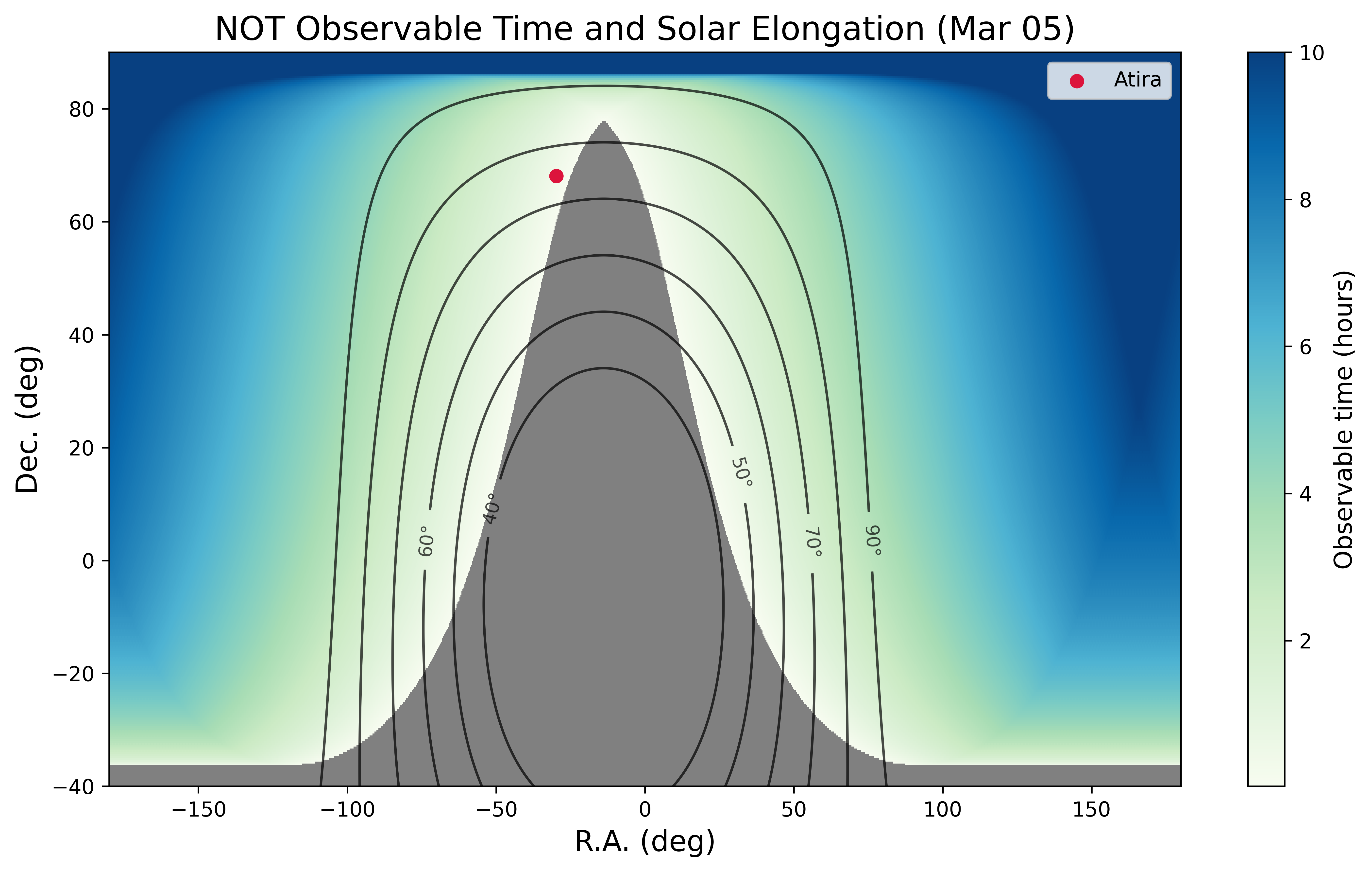}
    \caption{Observability for Metsähovi and NOT on March 05, showing the amount of time that a given sky location is above an elevation of $20\deg$ for Sun elevations less than $-18\deg$. Atira's 2024 sky position is given by a red dot and contour lines show the solar elongation of Solar System objects. The gray area signifies inaccessible sky locations for each observatory on this date.}
    \label{fig:observability}
\end{figure}

High-latitude observatories experience longer twilight and are thus uniquely positioned to observe asteroids interior to Earth. This includes the Atira class, so-called Vatiras (objects orbiting within the perihelion of Venus) and imminent impactors that come from the direction of the Sun. Depending on the ecliptic latitude of the target, observations conducted from locations above latitude $\sim50\deg$ can offer follow-up confirmation and physical characterisation of asteroids over several hours that would otherwise be observable from the equator for shorter periods during twilight. However, this location does not work as well for objects with low orbital inclinations, which are much less likely to appear at high enough declinations. Similarly, Venus co-orbital objects and the Vatiras that, at most, only reach solar elongations of 47$\deg$ \citep{Ye_etal20} may only be observed during twilight hours, regardless of the location on Earth. Thermal infrared observations provide a direct way to estimate the sizes of these populations and NEOSurveyor is designed to discover asteroids at low solar elongations. However, it will not acquire any optical brightness measurements needed to compute albedo or colors. 

Finally, we highlight the indifference of radar facilities with regard to the visible sky brightness. As mentioned above, near-surface properties can be ascertained from radar observations that are useful for studying surface properties. Further radar observations of Atira will not be possible in the next decade. Ideally, a radar facility that builds upon the legacy of Arecibo Observatory would be needed \citep{Venditti_etal23}.

\section{Conclusions \& Future Work}

Using color photometry and a visible spectrum we determined a Cgh-type classification in the SMASS taxonomy for Atira. The $0.7\um$ phyllosilicate feature is deeper and with a band position centered at longer wavelengths than any other reported in the PRIMASS Main Belt survey. Higher-quality visible spectra would better characterise the $0.7 \um$ phyllosilicate feature. Follow-up spectral observations at near-infrared wavelengths can better constrain the taxonomy, albedo, and regolith properties like grain size \citep{Vernazza_etal16}. Observing Atira's reflectance spectrum into the near-infrared region will give further insight into its primitive nature by placing further constrains on a meteorite analog. Given its proximity to the Sun, the thermal excess tail should be present and can be used to estimate the albedo. Given the high correlation with the 0.7$\um$ phyllosilicate feature \citep{Rivkin_etal15b} we expect Atira to exhibit a strong 3-$\um$ hydration feature.

Based on the degree of polarisation in comparison to other NEAs and its spectral classification, we infer that Atira's albedo lies in the range 0.05 -- 0.11. The previous albedo estimates of around 0.02 was derived from the relationship with the size and absolute magnitude. Since it is virtually impossible to observe Atira at phase angles less than $\sim70\deg$ \citep{rondon2022} any absolute magnitude estimation will be very uncertain. The albedo may be estimated via modeling of the thermal excess from spectroscopic observations \citep{Rivkin_etal05}. The size of Atira, as determined from delay-Doppler images, is also somewhat uncertain. Future ground-based observations of the thermal emission at mid-infrared wavelengths (e.g., the N-band centered around 10 $\um$) can be used to directly determine the size and used to improve on the bulk porosity estimated in our work of 51 $\pm$ 31\%. Thermal infrared observations should also provide insight into the thermal inertia that can be used to model regolith grain size \citep{Delbo_etal15,MacLennan+Emery21}.

Follow-up lightcurve, spectroscopic, polarimetric observations of Atira will contribute to more detailed knowledge of its physical and compositional properties. As of writing this article, an opportunity to observe Atira again from the NOT in the early part of 2026 has passed. The next best opportunity to observe Atira from Metsähovi would be in March-April 2031, when the brightness is better than 18 mag and the northern declination again reaches high values exceeding $60\deg$. Before this time, the brightness will stay fainter than $18\magnitude$, so would not be ideal for polarimetric or spectroscopic study with the NOT. Unfortunately, larger aperture telescopes like the VLT may not have the capability to access low elevations needed to target this asteroid.


\section*{Acknowledgements}

E.M.M. recognizes support from Research Council of Finland grant \#353784. Part of this work features observations conducted with the Nordic Optical Telescope (NOT), owned in collaboration by the University of Turku and Aarhus University, and operated jointly by Aarhus University, the University of Turku and the University of Oslo, representing Denmark, Finland and Norway, the University of Iceland and Stockholm University at the Observatorio del Roque de los Muchachos, La Palma, Spain, of the Instituto de Astrofisica de Canarias. The data presented here were obtained with ALFOSC, which is provided by the Instituto de Astrofisica de Andalucia (IAA) under a joint agreement with the University of Copenhagen and NOT. The NOT polarimetric and spectroscopic observations were carried out as part of an ESA Space Situational Awareness Program (S2P). The meteorite spectra shown herein were taken at the NASA RELAB facility at Brown University.

\section*{Data Availability}

The Metsähovi lightcurve photometry underlying this article is provided in the Appendix. The raw FITS files from the Nordic Optical Telescope (NOT) polarimetric and spectroscopic observations are available from the corresponding author upon reasonable request.



\bibliographystyle{mnras}
\bibliography{bibliography.bib}



\appendix

\section{Lightcurve Data}\label{app:A}

\begin{table}
\centering
\begin{tabular}{ccc}
\hline
JD & Magnitude & Uncertainty \\
\hline
2460375.3769676 & 17.196 & 0.109 \\
2460375.3826968 & 17.213 & 0.102 \\
2460375.3857986 & 17.369 & 0.108 \\
2460375.3889005 & 17.271 & 0.102 \\
2460375.3920023 & 17.251 & 0.079 \\
2460375.3951042 & 17.199 & 0.072 \\
2460375.4075116 & 17.255 & 0.076 \\
2460375.4172917 & 17.148 & 0.072 \\
2460375.4265972 & 17.216 & 0.072 \\
2460375.4328009 & 17.156 & 0.073 \\
2460375.4359028 & 17.198 & 0.069 \\
2460375.4390046 & 17.185 & 0.068 \\
2460375.4421065 & 17.170 & 0.068 \\
2460375.4452083 & 17.288 & 0.079 \\
2460375.4539931 & 17.154 & 0.067 \\
2460375.4570949 & 17.107 & 0.070 \\
2460375.4601968 & 17.287 & 0.074 \\
2460375.4880671 & 17.139 & 0.064 \\
2460375.4911690 & 17.232 & 0.067 \\
2460375.4942708 & 17.234 & 0.075 \\
2460375.4973727 & 17.182 & 0.069 \\
2460375.5004745 & 17.206 & 0.066 \\
2460375.5035764 & 17.090 & 0.064 \\
2460375.5097801 & 16.966 & 0.051 \\
2460375.5128819 & 16.996 & 0.051 \\
2460375.5159722 & 16.879 & 0.048 \\
2460375.5192708 & 17.021 & 0.057 \\
2460375.5223727 & 17.154 & 0.058 \\
2460375.5254745 & 17.078 & 0.056 \\
2460375.5285764 & 17.133 & 0.060 \\
2460375.5378819 & 17.196 & 0.059 \\
2460375.5409838 & 17.209 & 0.057 \\
2460375.5440856 & 17.188 & 0.061 \\
2460375.5471875 & 17.205 & 0.056 \\
2460375.5555093 & 17.149 & 0.059 \\
2460375.5586111 & 17.183 & 0.064 \\
2460375.5617014 & 17.161 & 0.059 \\
2460375.5648032 & 17.178 & 0.059 \\
2460375.5679051 & 17.158 & 0.062 \\
2460375.5710069 & 17.142 & 0.059 \\
2460375.5741088 & 17.149 & 0.057 \\
2460375.5803125 & 17.111 & 0.054 \\
2460375.5837153 & 17.235 & 0.060 \\
2460375.5868171 & 17.252 & 0.061 \\
2460375.5899190 & 17.249 & 0.066 \\
2460375.5930208 & 17.171 & 0.060 \\
2460375.5961111 & 17.129 & 0.057 \\
2460375.6023148 & 17.193 & 0.058 \\
2460375.6149537 & 17.018 & 0.051 \\
2460375.2945023 & 17.055 & 0.053 \\
2460375.3020602 & 17.110 & 0.057 \\
2460375.3096296 & 16.973 & 0.052 \\
2460375.3156829 & 16.901 & 0.050 \\
2460375.3388657 & 16.985 & 0.052 \\
2460375.3403704 & 16.992 & 0.053 \\
2460375.3523611 & 16.967 & 0.047 \\
2460375.3545718 & 16.971 & 0.056 \\
2460375.3560880 & 16.882 & 0.052 \\
2460375.3576042 & 17.063 & 0.054 \\
2460375.3591088 & 17.199 & 0.066 \\
2460375.3606250 & 17.272 & 0.067 \\
2460375.3621412 & 17.202 & 0.068 \\
2460375.3636458 & 17.166 & 0.068 \\
2460375.3651620 & 17.167 & 0.066 \\
2460375.3666782 & 17.241 & 0.067 \\
2460375.3681829 & 17.122 & 0.075 \\
\hline
\end{tabular}
\caption{Observed $V$ band Brightness}\label{tab:Vmags}
\end{table}

\begin{table}
\centering
\begin{tabular}{ccc}
\hline
JD & Magnitude & Uncertainty \\
\hline
2460375.3199653 & 16.902 & 0.072 \\
2460375.3214815 & 16.888 & 0.068 \\
2460375.3229977 & 16.906 & 0.066 \\
2460375.3245023 & 17.020 & 0.075 \\
2460375.3260185 & 17.037 & 0.079 \\
2460375.3275347 & 16.906 & 0.066 \\
2460375.3290394 & 16.995 & 0.074 \\
2460375.3305556 & 16.900 & 0.070 \\
2460375.3754398 & 16.739 & 0.085 \\
2460375.3811690 & 16.850 & 0.092 \\
2460375.3842708 & 16.755 & 0.083 \\
2460375.3873727 & 16.648 & 0.074 \\
2460375.3904745 & 16.818 & 0.070 \\
2460375.3935764 & 16.638 & 0.063 \\
2460375.4028819 & 16.499 & 0.050 \\
2460375.4059838 & 16.530 & 0.053 \\
2460375.4157639 & 16.501 & 0.045 \\
2460375.4188657 & 16.585 & 0.046 \\
2460375.4250694 & 16.642 & 0.051 \\
2460375.4312731 & 16.517 & 0.044 \\
2460375.4343750 & 16.675 & 0.049 \\
2460375.4374769 & 16.606 & 0.044 \\
2460375.4405787 & 16.799 & 0.055 \\
2460375.4436806 & 16.710 & 0.050 \\
2460375.4493634 & 16.791 & 0.058 \\
2460375.4524653 & 16.825 & 0.054 \\
2460375.4555671 & 16.834 & 0.057 \\
2460375.4586690 & 16.926 & 0.064 \\
2460375.4617708 & 16.719 & 0.051 \\
2460375.4648727 & 16.767 & 0.053 \\
2460375.4865394 & 16.801 & 0.070 \\
2460375.4896412 & 16.670 & 0.058 \\
2460375.4927431 & 16.620 & 0.060 \\
2460375.4958449 & 16.653 & 0.054 \\
2460375.4989468 & 16.744 & 0.062 \\
2460375.5020486 & 16.759 & 0.059 \\
2460375.5051505 & 16.740 & 0.058 \\
2460375.5082523 & 16.785 & 0.055 \\
2460375.5113542 & 16.751 & 0.054 \\
2460375.5144560 & 16.856 & 0.063 \\
2460375.5177431 & 16.825 & 0.061 \\
2460375.5208449 & 16.690 & 0.055 \\
2460375.5239468 & 16.779 & 0.056 \\
2460375.5270486 & 16.766 & 0.056 \\
2460375.5301505 & 16.693 & 0.053 \\
2460375.5363542 & 16.685 & 0.056 \\
2460375.5394560 & 16.576 & 0.048 \\
2460375.5425579 & 16.603 & 0.052 \\
2460375.5456597 & 16.678 & 0.057 \\
2460375.5508796 & 16.517 & 0.054 \\
2460375.5539815 & 16.610 & 0.055 \\
2460375.5570833 & 16.599 & 0.054 \\
2460375.5601852 & 16.569 & 0.050 \\
2460375.5632870 & 16.575 & 0.046 \\
2460375.5663773 & 16.604 & 0.052 \\
2460375.5694792 & 16.490 & 0.052 \\
2460375.5725810 & 16.580 & 0.056 \\
2460375.5756829 & 16.511 & 0.046 \\
2460375.5821875 & 16.699 & 0.058 \\
2460375.5852894 & 16.675 & 0.053 \\
2460375.5914931 & 16.692 & 0.060 \\
2460375.6007870 & 16.746 & 0.063 \\
2460375.6038889 & 16.731 & 0.057 \\
2460375.6069907 & 16.657 & 0.064 \\
2460375.6100810 & 16.729 & 0.074 \\
2460375.6134259 & 16.733 & 0.077 \\
\hline
\end{tabular}
\caption{Observed $R_C$ band Brightness}\label{tab:Rmags}
\end{table}


\bsp	
\end{document}